\documentclass[a4paper,rmp]{revtex4}

\usepackage{amsfonts}
\usepackage{amsthm}
\usepackage{amsmath}
\makeatletter \renewcommand\@biblabel[1]{$^{#1}$} \makeatother

\newtheorem{lemma}{Lemma}
\newtheorem{proposition}{Proposition}
\newtheorem{theorem}{Theorem}

\theoremstyle{remark}

\theoremstyle{definition}
\newtheorem{definition}{Definition}

\begin{document}

\title{On the mode stability of a self-similar wave map}

\author{Roland Donninger}
\email{roland.donninger@univie.ac.at}
\author{Peter C. Aichelburg}
\email{aichelp8@univie.ac.at}
\affiliation{Faculty of Physics, Gravitational Physics,  
University of Vienna,  
Boltzmanngasse 5, A-1090 Wien}


\begin{abstract}
We study linear perturbations of a self-similar wave map from
Minkowski space to the three-sphere which
is conjectured to be linearly stable.
Considering analytic mode solutions of the evolution equation for 
the perturbations
we prove that there are no real unstable eigenvalues apart from the well-known
gauge instability.
\end{abstract}

\maketitle

\section{Introduction}

\subsection{Motivation}
When considering the Cauchy problem for nonlinear evolution equations one often
encounters a behaviour known as "blow up": Evolutions with regular
initial data become singular after a finite time.
Moreover, it depends on the "size" of the data whether the blow up occurs or
not.
Such phenomena have been observed for many different equations arising from
various branches of physics, chemistry, biology, etc.
In particular Einstein's equations of general relativity have this property.
Furthermore, in many cases there is some kind of universal behaviour in the sense
that the shape of the blow up profile is independent of the special form of the
data.  

Co-rotational wave maps from
Minkowski space to the three-sphere have been 
used as a toy-model for blow up phenomena in general relativity 
(cf. Ref. \onlinecite{Bizon2000a}, Ref. \onlinecite{Bizon2000}).
Singularity formation for these wave maps has been studied extensively using 
numerical techniques but there are very few rigorous results available.
In numerical simulations (Ref. \onlinecite{Bizon2000}) one observes that singularity formation takes
place in a universal manner via a certain self-similar solution $f_0$.
The existence of $f_0$ has been established rigorously in Ref. \onlinecite{Shatah1988} and
later it has been found in closed form (Ref. \onlinecite{Turok1990}).
Using adapted coordinates, this self--similar blow up can be reformulated as an 
asymptotic stability problem.
Thus, it is conjectured that the solution $f_0$ is linearly stable.

However, the eigenvalue equation 
for perturbation modes around $f_0$ (although a linear
ordinary differential equation of second order) 
is hard to handle.
In particular, it is not possible to analyse it with standard self--adjoint 
Sturm--Liouville techniques.
Therefore, not even mode stability (i.e. the non-existence of eigenvalues 
with positive real parts) 
of $f_0$ has been proved so far.
Nevertheless, Sturm--Liouville theory can be applied at least partly and
Bizo\'n was able 
to 
show that there are no eigenvalues with real parts greater than 1 
(Ref. \onlinecite{Bizon2005}).  
The main result of the present paper is the non--existence of mode
solutions with eigenvalues between $0$ and $1$ which further supports the
conjecture of linear stability of $f_0$.

\subsection{Derivation of the equations, main results}
We study wave maps from Minkowski space to the
three-sphere (see Ref. \onlinecite{Bizon2000a} for a definition) and
restrict ourselves to co-rotational solutions which reduces
the problem to the single semilinear wave equation
\begin{equation}
\label{semlin}
\Psi_{tt}(t,r)=\Psi_{rr}(t,r)+\frac{2}{r}\Psi_r(t,r)-\frac{\sin(2
\Psi(t,r))}{r^2}.
\end{equation}
Since we are interested in self-similar solutions we introduce adapted
coordinates $\tau:=-\log(T-t)$ and $\rho:=r/(T-t)$ where $T>0$ is an arbitrary
constant (the blow up time).
The self--similar solution $f_0$ is given by 
$f_0(\rho)=2\arctan(\rho)$.
We consider linear perturbations of $f_0$ and insert the ansatz
$$ \Psi(\tau, \rho)=f_0(\rho)+w(\tau,\rho) $$
into eq. (\ref{semlin}).
Neglecting higher-order terms leads to
\begin{equation}
\label{pert}
w_{\tau \tau}-(1-\rho^2)w_{\rho \rho}+2\rho w_{\tau \rho}+w_\tau -2
\frac{1-\rho^2}{\rho}w_\rho+\frac{2\cos(2 f_0)}{\rho^2}w=0
\end{equation}
which is a time evolution equation for linear perturbations $w$ of $f_0$.
The wave map $f_0$ is considered to be linearly stable if all solutions of eq.
(\ref{pert}) remain bounded (with respect to a suitable norm) for all $\tau>0$.

We are interested in mode solutions of eq. (\ref{pert}), i.e. solutions of the
form 
\begin{equation*}
\label{mode}
w(\tau,\rho)=e^{\lambda \tau}u(\rho) 
\end{equation*}
for $\lambda \in \mathbb{C}$.
Inserting this ansatz into eq. (\ref{pert}) yields 
\begin{equation}
\label{pencil}
u''+\left ( \frac{2}{\rho}-\frac{2 \lambda \rho}{1-\rho^2} \right )
u'-\left ( 
\frac{2 \cos(2 f_0)}{\rho^2 (1-\rho^2)}+
\frac{\lambda(1+\lambda)}{1-\rho^2} \right ) u=0
\end{equation}
which is a second-order ordinary differential equation (ODE) for $u$ with singular
points at $\rho=0$ and $\rho=1$.
Thus, a necessary condition for linear stability of $f_0$ is the non-existence
of mode solutions with $\mathrm{Re}\lambda>0$.
It is a subtle question what degree of regularity one should require to consider
a solution of eq. (\ref{pencil}) as "admissible".
However, we will at least assume that any admissible solution should be
differentiable on $[0,1]$.
Thus, by a regular solution of eq. (\ref{pencil}) we mean a continuous 
function $u: [0,1]
\to \mathbb{C}$ that solves eq. (\ref{pencil}) on $(0,1)$ in the classical sense
and $u'$ can be continuously extended to the whole interval $[0,1]$. 
We remark that eq. (\ref{pencil}) does not constitute a standard eigenvalue 
problem since 
the coefficient of
the first derivative of $u$ depends on the spectral parameter $\lambda$.
Nevertheless we will speak of eigenvalues and eigenfunctions.
A simple change of the dependent variable $u \mapsto \widetilde{u}$ where 
$$ \tilde{u}(\rho):=\rho (1-\rho^2)^{\lambda/2} u(\rho) $$
leads to the equation
\begin{equation}
\label{sl}
\tilde{u}''-\left ( \frac{2 \cos(2f_0)}{\rho^2
(1-\rho^2)}-\frac{\lambda(2-\lambda)}{(1-\rho^2)^2} \right ) \tilde{u}=0.
\end{equation}
Both forms eq. (\ref{pencil}) and eq. (\ref{sl}) of the eigenvalue problem will
be useful in the sequel.
Now we are ready to formulate our result.

\begin{theorem}
\label{nonexistence}
For $\lambda \in (0,1)$ there does
not exist a regular solution of eq. (\ref{pencil}).
\end{theorem}

\subsection{Notations}
We will make frequent use of the following standard notations.
By $C^k[a,b]$ we denote the vector space of continuous functions 
$u: [a,b] \to \mathbb{C}$ 
which are
$k$--times continuously differentiable on $(a,b)$ and all $k$ derivatives
can be extended to continuous functions on $[a,b]$.
On this space we define a norm by
$$ \|u\|_{C^k[a,b]}:=\sum_{j=0}^k \sup_{x \in [a,b]}|u^{(j)}(x)|. $$
The normed vector space $C^k[a,b]$ together with $\|\cdot\|_{C^k[a,b]}$ is a
Banach space.
We also mention (weighted) Lebesgue spaces 
$$ L^p((a,b), w(x)dx) $$ defined by (equivalence classes
of) functions $u: [a,b] \to \mathbb{C}$ such that
$$ \int_a^b |u(x)|^p w(x)dx < \infty $$
where the integral is understood in the sense of Lebesgue.

\section{Properties of the eigenvalue equation, known results}
\subsection{The gauge mode}
For $\lambda=1$ there is an analytic solution of eq. (\ref{pencil}) which can be
given in closed form, the so--called gauge mode (cf. Ref. \onlinecite{Bizon2000}) 
\begin{equation}
\label{gauge}
\theta(\rho):=\frac{2 \rho}{1+\rho^2}.
\end{equation}
This gauge mode leads to an exponentially growing solution 
$w(\tau, \rho)=e^\tau \theta(\rho)$ 
of eq. (\ref{pert}) which seems to spoil linear stability of $f_0$.
However, this instability is connected to the freedom of 
choosing the blow up time $T$
when introducing the adapted coordinates $(\tau,\rho)$ (cf. Ref. \onlinecite{Bizon2000}).
Therefore, we are only interested in stability modulo this gauge freedom.
Nevertheless, the existence of this gauge mode will play an essential role in 
the further analysis. 

\subsection{Asymptotic estimates}
Eq. (\ref{pencil}) has regular singular points at $\rho=0, 1$.
Using Frobenius' method we can derive asymptotic estimates for solutions of eq.
(\ref{pencil}).
Around $\rho=0$ there exists a regular solution $\varphi_0$ and a singular
solution $\psi_0$. Around $\rho=1$ the situation is similar but more
subtle (cf. table \ref{asymest}, $c$ is a constant which might also be zero).
\begin{table}[h]
\begin{center}
\begin{tabular}{|l|l|l|l|}
\hline
$\rho$ & $\lambda$ & Analytic solution & Non-analytic solution \\
\hline
$\rho \rightarrow 0$ & any & $\varphi_0 \sim \rho$ & 
$\psi_0 \sim \rho^{-2}$ \\
\hline
$\rho \rightarrow 1$ & $\lambda \notin \mathbb{Z}$ & $\varphi_1 
\sim 1$ & $\psi_1 \sim (1- \rho)^{1- \lambda}$ \\
& $\lambda \in \mathbb{Z}$, $\lambda > 1$ & $\varphi_1 \sim 1$ &
$\psi_1 \sim c\log (1- \rho)+(1- \rho)^{1- \lambda}$ \\
& $\lambda \in \mathbb{Z}$, $\lambda \leq 1$ & $\varphi_1 \sim 
(1- \rho)^{1- \lambda}$ & $\psi_1 \sim c (1- \rho)^{1-
\lambda}\log (1- \rho) +1$ \\
\hline
\end{tabular}
\end{center}
\caption{Asymptotic estimates for solutions of eq. (\ref{pencil})}
\label{asymest}
\end{table} 

\begin{definition}
For a given $\lambda \in \mathbb{C}$ we denote by $\varphi_0(\cdot, \lambda)$
the solution of eq. (\ref{pencil}) which is analytic around $\rho=0$ and
satisfies $\varphi_0'(0,\lambda)=2$ ($'$ denotes $d/d\rho$). 
Similarly, by $\varphi_1(\cdot, \lambda)$ we denote the solution of eq.
(\ref{pencil}) which is analytic around $\rho=1$ and satisfies
$\varphi_1(1,\lambda)=1$.
\end{definition} 

If the dependence on $\lambda$ is not essential we will 
sometimes omit it in the argument. For instance we will occasionally write 
$f$ instead of $f(\cdot, \lambda)$ or $f(\rho)$ instead of $f(\rho, \lambda)$
for a function $f$ of $\rho$ and $\lambda$.

Due to the location of the singularities we conclude that 
$\varphi_0$ and  $\varphi_1$ are analytic on $[0,1)$ and $(0,1]$, respectively. 
For $\rho \in (0,1)$, eq. (\ref{pencil}) is perfectly regular (all coefficients
are analytic) and well--known theorems on linear ODEs tell us that the 
solution space is two--dimensional.
Since $\varphi_1$ and $\psi_1$ are linearly independent, it follows that there
exist constants $c_1$ and $c_2$ such that $\varphi_0=c_1\varphi_1 + c_2\psi_1$ on
$(0,1)$. 
If $c_2=0$ then $\varphi_0$ is analytic on $[0,1]$ and it is therefore an analytic
eigenfunction.
Moreover, as long as $\mathrm{Re} \lambda>0$, the analytic eigenfunctions are
exactly the regular ones we are interested in.

\subsection{Sturm--Liouville theory}
At first glance one might expect 
Sturm--Liouville theory to answer all questions
concerning solutions of eq. (\ref{pencil}) since it can be transformed to
Sturm--Liouville form (eq. (\ref{sl})).
Unfortunately this is not true. 
It turns out that the differential operator defined by eq. (\ref{sl}) is
symmetric on the weighted Lebesgue space 
$$ L^2\left((0,1), \frac{d\rho}{(1-\rho^2)^2}\right) $$
but for $\mathrm{Re}\lambda \leq 1$ the solution 
$\tilde{\varphi}_1(\rho, \lambda):=\rho (1-\rho^2)^{\lambda/2}\varphi_1(\rho,
\lambda)$ we are interested in is not an element of this space.
Therefore, Sturm--Liouville theory is useless for $\mathrm{Re}\lambda \leq 1$.
Nevertheless, for $\mathrm{Re}\lambda > 1$ it can be applied and it follows
that there do not exist analytic eigenfunctions for $\mathrm{Re}\lambda > 1$.
 
\section{Integral equations}
In what follows we derive integral 
equations for $\varphi_0$ and $\varphi_1$. 
This is straight--forward although sometimes a little tricky.

\subsection{An integral equation for $\varphi_0$}
We split eq. (\ref{pencil}) as
\begin{equation}
\label{inhomog}
u''+\left (\frac{2}{\rho}-\frac{2 \rho}{1-\rho^2} \right )
u'-\left ( \frac{2 \cos(f_0)}{\rho^2(1-\rho^2)}+\frac{2}{1-\rho^2}
\right ) u=Q_\lambda u
\end{equation}
where
$$ Q_\lambda u(\rho, \lambda):=\frac{1}{1-\rho^2}\left \{ 2(\lambda-1)\rho 
u'(\rho, \lambda) + [\lambda(1+\lambda)-2]u(\rho, \lambda) \right\}. $$

The idea is to interpret eq. (\ref{inhomog}) as an "inhomogeneous" equation and
apply the variations of constants formula.
First of all we need a fundamental system for the "homogeneous"
equation
\begin{equation}
\label{homog}
u''+\left (\frac{2}{\rho}-\frac{2 \rho}{1-\rho^2} \right )
u'-\left ( \frac{2 \cos(f_0)}{\rho^2(1-\rho^2)}+\frac{2}{1-\rho^2}
\right ) u=0
\end{equation}
Since eq. (\ref{homog}) is exactly eq. (\ref{pencil}) with $\lambda=1$ it is
clear that the gauge mode $\theta$ solves eq. (\ref{homog}).
Another linearly independent solution is
$$ \chi(\rho):=\frac{1}{1+\rho^2}\left (\frac{1}{\rho^2}+6 \rho \log \left
(\frac{1-\rho}{1+\rho}\right )+9 \right ). $$
Therefore $\{\theta,\chi\}$ is a fundamental system for eq. (\ref{homog}).
The Wronskian $W(\theta, \chi):=\theta \chi' - \theta' \chi$ of $\theta$ and 
$\chi$ is given by
$$ W(\theta, \chi)(\rho)=-\frac{6}{\rho^2 (1-\rho^2)}. $$
Inspired by the variations of constants formula we consider the integral equation
\begin{equation}
\label{inteqphi0}
\begin{split}
u(\rho, \lambda)=\theta(\rho)-\theta(\rho)\int_0^\rho
\frac{\chi(\xi)}{W(\theta,\chi)(\xi)}Q_\lambda u(\xi,\lambda)d\xi \\
+\chi(\rho)\int_0^\rho 
\frac{\theta(\xi)}{W(\theta,\chi)(\xi)}Q_\lambda u(\xi,\lambda)d\xi
\end{split}
\end{equation}
for $\rho \in [0,1)$.

First we will show that studying eq. (\ref{inteqphi0}) tells us something
about solutions of eq. (\ref{pencil}).

\begin{lemma}
\label{leminteqdiff0}
For some $\rho_0 \in (0,1)$ and a given $\lambda \in \mathbb{C}$ let $u \in
C^1[0,\rho_0]$ be a solution of eq. (\ref{inteqphi0}).
Then $u \in C^2[0,\rho_0]$ and $u$ solves eq. (\ref{pencil}).
\end{lemma}

\begin{proof}
For $\rho \in (0,\rho_0]$ the right-hand side of eq. (\ref{inteqphi0}) is
obviously twice continuously differentiable.
By defining $u''(0):=0$ we obtain $u \in C^2[0,\rho_0]$ which is then 
by construction (variations of constants formula) also a
solution of eq. (\ref{pencil}).
\end{proof} 

In what follows we show that eq. (\ref{inteqphi0}) has a solution $u$ for any 
$\lambda \in \mathbb{C}$. 
This solution $u$ is also a solution of eq. (\ref{pencil}) and satisfies
$u(0,\lambda)=0$ as well as $u'(0,\lambda)=2$. 
Therefore we have $u=\varphi_0$ and eq. (\ref{inteqphi0}) is an integral
equation for $\varphi_0$.
The key ingredient is the following Proposition the proof of which
will be postponed to appendix \ref{proofexintphi0}.

\begin{proposition}
\label{exintphi0}
Let $\lambda \in \mathbb{C}$.
Then there exists a $\rho_0 \in (0,1)$ such
that the integral equation
(\ref{inteqphi0}) has a solution $u(\cdot,\lambda) 
\in C^1[0,\rho_0]$.
\end{proposition}

\begin{proof}
See appendix \ref{proofexintphi0}.
\end{proof}

\begin{lemma}
\label{lemphi0int}
Let $\lambda \in \mathbb{C}$. Then there exists a $\rho_0 \in (0,1)$ such that
$\varphi_0(\cdot, \lambda)$ solves eq. 
(\ref{inteqphi0}) on $[0, \rho_0]$.
\end{lemma}

\begin{proof}
From Proposition \ref{exintphi0} we know that there exists a $\rho_0 \in (0,1)$
such that there is a solution $u(\cdot, \lambda) \in C^1[0,\rho_0]$ of eq. 
(\ref{inteqphi0}).
Using Lemma \ref{leminteqdiff0} we conclude that $u(\cdot, \lambda)$ is in 
fact twice continuously
differentiable and solves eq. (\ref{pencil}).
From eq. (\ref{inteqphi0}) we observe that $u
\not= 0$, $u(0,\lambda)=0$, $u'(0,\lambda)=2$.
On $(0,1)$ the space of solutions of eq. (\ref{pencil}) is spanned by $\varphi_0$
and $\psi_0$ where $\psi_0$ is a solution of eq. (\ref{pencil}) which is
singular at $\rho=0$. Therefore we have $u=c_1\varphi_0 + c_2\psi_0$ for constants
$c_1$ and $c_2$ on $(0,1)$.
The conditions $\lim_{\rho \to 0}u(\rho, \lambda)=0$ and 
$\lim_{\rho \to 0}u'(\rho, \lambda)=2$ yield $c_2=0$ and
$c_1=1$ and thus $\varphi_0=u$.

\end{proof}   

As a last step we claim that $\varphi_0$ actually satisfies the integral
equation (\ref{inteqphi0}) on any $[0,a] \subset [0,1)$ and not 
only on $[0,\rho_0]$ for the special $\rho_0$ from Lemma \ref{lemphi0int}.

\begin{proposition}
\label{propphi0int}
Let $[0, a] \subset [0,1)$ and $\lambda \in \mathbb{C}$. 
Then the function $\varphi_0(\cdot, \lambda)$ satisfies eq. (\ref{inteqphi0})
for all $\rho \in [0,a]$.
\end{proposition}

\begin{proof}
Let $\rho_0 \in (0,1)$ be the constant from 
Lemma \ref{lemphi0int} and choose $a \in (\rho_0,1)$.
Since eq. (\ref{pencil}) is regular on $[\rho_0,a]$ everything follows from
well-known ODE theory.
\end{proof}

\subsection{An integral equation for $\varphi_1$}
We deduce a similar integral equation for $\varphi_1$. However, due to the
singularity of $Q_\lambda u$ at $\rho=1$ we have to split eq. (\ref{pencil})
in a different way. We write 
$$ u''+\left (\frac{2}{\rho}-\frac{2 \lambda \rho}{1-\rho^2} \right )u'=qu $$
where
$$ q(\rho, \lambda):=\frac{\lambda(1+\lambda)}{1-\rho^2}+\frac{2 \cos(2
f_0)}{\rho^2(1-\rho^2)}. $$
$\{1, \psi\}$ is a fundamental system for the "homogeneous" equation where 
$$ \psi(\rho, \lambda):=\int_c^\rho \frac{d\xi}{\xi^2
(1-\xi^2)^\lambda} $$
and $c \in (0,1)$ is arbitrary.
The Wronskian is simply given by $W(1, \psi)=-\psi'$.
Thus we consider the integral equation
\begin{equation}
\label{inteqphi1}
\begin{split}
u(\rho, \lambda)=1-\int_\rho^1 \frac{\psi(\xi,\lambda)}{\psi'(\xi,\lambda)}q(\xi,
\lambda)u(\xi, \lambda)d\xi \\
+\psi(\rho, \lambda)\int_\rho^1 \frac{1}{\psi'(\xi, \lambda)}q(\xi,
\lambda)u(\xi,\lambda)d\xi
\end{split}  
\end{equation}
for $\rho \in (0,1]$.
To ensure existence of the integrals we have to restrict ourselves to
$\mathrm{Re}\lambda>0$ which will be assumed from now on.

Now we proceed similarly to the last section with the sole difference that it is
sufficient to consider continuous solutions of eq. (\ref{inteqphi1}) since no
derivative of $u$ appears in eq. (\ref{inteqphi1}).
\begin{lemma}
\label{leminteqdiff1}
For some $\rho_1 \in (0,1)$ and a given $\lambda \in \mathbb{C}$ with
$\mathrm{Re}\lambda > 0$ let $u \in
C[\rho_1,1]$ be a solution of eq. (\ref{inteqphi1}).
Then $u \in C^2[\rho_1,1]$ and $u$ solves eq. (\ref{pencil}).
\end{lemma}

\begin{proof}
Similar to the proof of Lemma \ref{leminteqdiff0}.
\end{proof}

\begin{proposition}
\label{exintphi1}
For a given $\lambda \in \mathbb{C}$ with $\mathrm{Re}\lambda > 0$ 
there exists a $\rho_1 \in (0,1)$ 
such that the integral equation
(\ref{inteqphi1}) has a solution 
$u(\cdot,\lambda) \in C[\rho_1,1]$.
\end{proposition}

\begin{proof}
See appendix \ref{proofexintphi1}.
\end{proof}

\begin{lemma}
\label{lemphi1int}
Let $\lambda \in \mathbb{C}$ with $\mathrm{Re}\lambda>0$. Then there exists a 
$\rho_1 \in (0,1)$ such that $\varphi_1(\cdot, \lambda)$ solves eq. 
(\ref{inteqphi1}) on $[\rho_1,1]$. 
\end{lemma}

\begin{proof}
Similar to the proof of Lemma \ref{lemphi0int}.
\end{proof}

\section{Properties of eigenfunctions}
We derive some properties of analytic eigenfunctions. 
We will show that (real) eigenfunctions do not have zeros on $(0,1)$ and that 
the derivative has to change sign on $(0,1)$ if $\lambda > -2$.
The first result is established by using a well--known method from classic
oscillation theory.

\begin{lemma}
\label{nozeros} 
For $\lambda \in \mathbb{R}$ let $u$ be a real solution of eq. 
(\ref{pencil}) on $[0,1)$ satisfying $u(0)=0$ and
$u'(0)>0$. Then $u(\rho)>0$ for all $\rho \in (0,1)$.
\end{lemma} 

\begin{proof}
Since the Lemma is obviously true for $\lambda=1$ we assume $\lambda \not= 1$.
We argue by contradiction. Suppose $u$ is a solution of eq. (\ref{pencil}) with
$u(0)=0$, $u'(0)>0$ and let $\rho_0 \in (0,1)$ be the smallest zero of $u$.
Thus we have $u(\rho)>0$ for all $\rho \in (0,\rho_0)$.
We define a new dependent variable $\tilde{u}$ by
$\tilde{u}(\rho):=\rho(1-\rho^2)^{\lambda/2}u(\rho)$.
Then we have $\tilde{u}(0)=0$, 
$\tilde{u}(\rho_0)=0$, $\tilde{u}(\rho)>0$ for all $\rho \in (0,\rho_0)$ 
and $\tilde{u}$ satisfies
the equation
\begin{equation}
\label{slp}
\tilde{u}''+q \tilde{u}=p_\lambda \tilde{u} 
\end{equation}
where
$$q(\rho)=-\frac{2 \cos(2 f_0)}{\rho^2 (1-\rho^2)}$$
and 
$$ p_\lambda(\rho)=-\frac{\lambda (2-\lambda)}{(1-\rho^2)^2}. $$ 
Observe that $p_\lambda-p_1 > 0$ for all $\lambda \in \mathbb{R}$, $\lambda
\not=1$.
Now let $\tilde{\theta}(\rho):=\rho \sqrt{(1-\rho^2)}\theta(\rho)$ where 
$\theta$ is the gauge mode (eq. \ref{gauge}).
Then $\tilde{\theta}$ satisfies eq. (\ref{slp}) with $\lambda=1$.  
Integration by parts yields
$$ \int_a^b \left [p_\lambda(\rho)-p_1(\rho) \right ]
\tilde{u}(\rho)\tilde{\theta}(\rho)d\rho
=W(\tilde{u},\tilde{\theta})(a)-W(\tilde{u},\tilde{\theta})(b) $$
for $a,b \in [0,1)$.
Now choose $a=0$ and $b=\rho_0$ to obtain
$$ \int_0^{\rho_0}
\left [ p_\lambda(\rho)-p_1(\rho) \right ] \tilde{u}(\rho)
\tilde{\theta}(\rho)d\rho=
\tilde{u}'(\rho_0)\tilde{\theta}(\rho_0). $$
But this is a contradiction since the left-hand side is positive while the
right-hand side is negative or equals zero ($\tilde{u}'(\rho_0) \leq 0$ since 
$\tilde{u}(\rho)>0$ for all $\rho \in (0,\rho_0)$).
\end{proof}

For the next result we make use of the fact that $\varphi_0$ is a solution of
eq. (\ref{inteqphi0}) on any $[0,a] \subset [0,1)$.
\begin{lemma}
\label{intid}
Let $u \in C^2[0,1]$ be a nontrivial solution of eq.
(\ref{pencil}) for some $\lambda \in \mathbb{C}$. Then $u$ satisfies 
$$ \int_0^1 \frac{\theta(\xi)}{W(\theta,\chi)(\xi)}Q_\lambda u(\xi,\lambda)
d\xi=0. $$  
\end{lemma}

\begin{proof}
For $\lambda=1$ the Lemma is trivial, so assume $\lambda \not=1$.
Since $u$ is regular around $\rho=0$ and not identically zero 
it follows that there exists a constant $c$ such
that $u=c\varphi_0$.
According to Proposition \ref{propphi0int}, $u$ satisfies
\begin{equation}
\label{inteqcu}
\begin{split}
u(\rho, \lambda)=c\theta(\rho)-\theta(\rho)\int_0^\rho
\frac{\chi(\xi)}{W(\theta,\chi)(\xi)}Q_\lambda u(\xi,\lambda)d\xi \\
+\chi(\rho)\int_0^\rho 
\frac{\theta(\xi)}{W(\theta,\chi)(\xi)}Q_\lambda u(\xi,\lambda)d\xi
\end{split}
\end{equation}  
for all $\rho \in [0,1)$.
Now suppose 
$$ \int_0^1 \frac{\theta(\xi)}{W(\theta,\chi)(\xi)}Q_\lambda u(\xi,\lambda)
d\xi\not=0. $$
Then passing to the limit $\rho \to 1$ in eq. (\ref{inteqcu}) yields divergence
of $u(\rho, \lambda)$ since $\chi$ diverges at $\rho=1$ and
$$ \frac{\chi}{W(\theta,\chi)}Q_\lambda u(\cdot,\lambda) \in L^1(0,1). $$
However, this is a contradiction to $u \in C^2[0,1]$.
\end{proof}

With these preparations we show that the derivative of regular solutions of eq.
(\ref{pencil}) has to change sign if $\lambda > -2$.

\begin{lemma}
\label{chsign}
For $\lambda > -2$, $\lambda \not= 1$ let $u \in C^2[0,1]$ be a nontrivial real solution of eq.
(\ref{pencil}). Then $u'$ changes its sign on $(0,1)$.
\end{lemma}

\begin{proof}
Suppose $u'$ does not change sign on $(0,1)$. Without loss of generality we
assume $u'(\rho) \geq 0$ for all $\rho \in [0,1)$. 
It follows that $u(\rho)>0$ for
all $\rho \in (0,1)$ ($u(0)=0$). By Lemma \ref{intid}, $u$ satisfies
\begin{equation}
\label{intidproof}
\int_0^1 \frac{\theta(\rho)}{W(\theta,\chi)(\rho)}
\frac{1}{1-\rho^2}\left \{ 2(\lambda-1)\rho u'(\rho) + 
[\lambda(1+\lambda)-2]u(\rho) \right \}d\rho=0. 
\end{equation}
But since $\mathrm{sgn}(\lambda -1)=\mathrm{sgn}(\lambda(1+\lambda)-2)$ 
the integrand in eq.
(\ref{intidproof}) is either positive or negative on $(0,1)$ which is
a contradiction.
\end{proof}

\section{Proof of the main result}
Now we are able to prove our main result.

\begin{proof}[Proof of Theorem \ref{nonexistence}]
Suppose $u \in C^2[0,1]$ is a nontrivial solution of eq. (\ref{pencil}).
Without loss of generality we assume $u$ to be real and $u'(0)>0$.
By Lemma \ref{nozeros} and the asymptotic estimates for $\varphi_1$ 
we observe that $u(1)>0$.
Furthermore, $u$ has to satisfy the regularity condition 
$$ u'(1)=\frac{2-\lambda - \lambda^2}{2 \lambda}u(1) $$
which follows from the existence of $u''(1)$.
This means that $u(1)$ and $u'(1)$ have the same sign ($\lambda \in (0,1)$)
and therefore $u'(1)>0$.
By Lemma \ref{chsign} we know that $u$ has a maximum on $(0,1)$, say at 
$\rho_0 \in (0,1)$.
We have $u(\rho_0)>0$, $u'(\rho_0)=0$ and $u''(\rho_0)<0$.
Inserting in eq. (\ref{pencil}) yields
$$ u''(\rho_0)=\beta_\lambda(\rho_0)u(\rho_0) < 0 $$
where 
$$ \beta_\lambda(\rho):=\frac{\lambda(1+\lambda)}{1-\rho^2}
+\frac{2 \cos(f_0(\rho))}{\rho^2(1-\rho^2)}. $$ 
Note that $\beta_\lambda$ is positive for small $\rho$ and has exactly one zero
on $(0,1)$,
say at $\rho_\lambda^* \in (0,1)$.
It follows that $\rho_0 > \rho_\lambda^*$.
Since $u'(1)>0$, $u'$ has to change sign again, say at $\rho_1 \in (\rho_0,1)$.
Therefore we have $u(\rho_1)>0$, $u'(\rho_1)=0$ and $u''(\rho_1)>0$.
Eq. (\ref{pencil}) yields
$$ u''(\rho_1)=\beta_\lambda(\rho_1)u(\rho_1)>0 $$
which is, however, impossible since $\beta_\lambda(\rho_1)<0$.
\end{proof}

\begin{acknowledgments}
We thank Piotr Bizo\'n for helpful discussions. 
This work was supported by the Austrian Fond zur  
F\"orderung der wissenschaftlichen Forschung (FWF) Project P19126.
\end{acknowledgments}


\begin{appendix}
\section{Existence of solutions of integral equations}

\subsection{Proof of Proposition \ref{exintphi0}}
\label{proofexintphi0}

\begin{proof}[Proof of Proposition \ref{exintphi0}]
Define the mapping $K: C^1[0,\rho_0] \to C^1[0,\rho_0]$ by
$$ Ku(\rho):=\theta(\rho)-\theta(\rho)\int_0^\rho
\frac{\chi(\xi)}{W(\theta,\chi)(\xi)}Q_\lambda u(\xi)d\xi 
+\chi(\rho)\int_0^\rho 
\frac{\theta(\xi)}{W(\theta,\chi)(\xi)}Q_\lambda u(\xi)d\xi
$$ 
where $\rho_0 \in (0,1)$ is to be chosen later.
Differentiation yields
\begin{equation*}
(Ku)'(\rho)=\theta'(\rho)-\theta'(\rho)\int_0^\rho
\frac{\chi(\xi)}{W(\theta,\chi)(\xi)}Q_\lambda u(\xi)d\xi 
+\chi'(\rho)\int_0^\rho 
\frac{\theta(\xi)}{W(\theta,\chi)(\xi)}Q_\lambda u(\xi)d\xi.
\end{equation*}
For $u,v \in C^1[0,\rho_0]$ we readily estimate
\begin{equation*}
\|Ku-Kv\|_{C^1[0,\rho_0]} \leq C \|\alpha+\beta\|_{C[0,\rho_0]} 
\|u-v\|_{C^1[0,\rho_0]}
\end{equation*}
where 
\begin{equation*}
\alpha(\rho):=|\theta(\rho)| \int_0^\rho
\frac{|\chi(\xi)|}{(1-\xi^2)|W(\theta,\chi)(\xi)|}d\xi
+|\chi(\rho)| \int_0^\rho
\frac{|\theta(\xi)|}{(1-\xi^2)|W(\theta,\chi)(\xi)|}d\xi
\end{equation*}
and 
\begin{equation*}
\beta(\rho):=|\theta'(\rho)| \int_0^\rho
\frac{|\chi(\xi)|}{(1-\xi^2)|W(\theta,\chi)(\xi)|}d\xi
+|\chi'(\rho)| \int_0^\rho
\frac{|\theta(\xi)|}{(1-\xi^2)|W(\theta,\chi)(\xi)|}d\xi
\end{equation*}
and $C>0$ is a constant.
Using de l'Hospital's rule one readily
shows that 
$\lim_{\rho \to 0}
(\alpha(\rho)+\beta(\rho))=0 $
and therefore we can choose $\rho_0>0$ so small that 
$ C':=C \|\alpha+\beta\|_{C[0,\rho_0]} <1 $ and $K$ satisfies the contraction
property. 
Hence, existence of a fixed point of $K$ follows from the contraction mapping
principle on the Banach space $C^1[0,\rho_0]$. 
\end{proof}

\subsection{Proof of Proposition \ref{exintphi1}}
\label{proofexintphi1}
\begin{proof}[Proof of Proposition \ref{exintphi1}]
For fixed $\lambda \in \mathbb{C}$ with $\mathrm{Re}\lambda>0$ we define the 
mapping $K: C[\rho_1,1] \to C[\rho_1,1]$ by
\begin{equation*}
Ku(\rho):=1-\int_\rho^1 \frac{\psi(\xi,\lambda)}{\psi'(\xi,\lambda)}q(\xi,
\lambda)u(\xi)d\xi 
+\psi(\rho, \lambda)\int_\rho^1 \frac{1}{\psi'(\xi, \lambda)}q(\xi,
\lambda)u(\xi)d\xi.
\end{equation*}
For $u,v \in C[\rho_1,1]$ we estimate
\begin{equation*}
\|Ku-Kv\|_{C[\rho_1,1]}  \leq 
\|\alpha(\cdot, \lambda)\|_{C[\rho_1,1]} \|u-v\|_{C[\rho_1,1]} 
\end{equation*}
where 
$$ \alpha(\rho, \lambda)=\int_\rho^1 \left |
\frac{\psi(\xi,\lambda)}{\psi'(\xi,\lambda)}q(\xi,
\lambda) \right | d\xi
+| \psi(\rho, \lambda) | \int_\rho^1 \left | \frac{q(\xi,
\lambda)}{\psi'(\xi, \lambda)} \right | d\xi. $$
De l'Hospital's rule implies
that $\lim_{\rho \to 1}\alpha(\rho,\lambda)=0$ and thus, for 
$\rho_1$ sufficiently close to $1$, the contraction mapping principle on
$C[\rho_1,1]$ guarantees the existence of a fixed point of $K$.
\end{proof}

\end{appendix}


\begin{thebibliography}{1}

\bibitem{Bizon2000a}
Piotr Bizo{\'n}.
\newblock Equivariant self-similar wave maps from {M}inkowski spacetime into
  3-sphere.
\newblock {\em Comm. Math. Phys.}, 215(1):45--56, 2000.

\bibitem{Bizon2005}
Piotr Bizo{\'n}.
\newblock An unusual eigenvalue problem.
\newblock {\em Acta Phys. Polon. B}, 36(1):5--15, 2005.

\bibitem{Bizon2000}
Piotr Bizo{\'n}, Tadeusz Chmaj, and Zbis{\l}aw Tabor.
\newblock Dispersion and collapse of wave maps.
\newblock {\em Nonlinearity}, 13(4):1411--1423, 2000.

\bibitem{Shatah1988}
Jalal Shatah.
\newblock Weak solutions and development of singularities of the {${\rm
  SU}(2)$} {$\sigma$}-model.
\newblock {\em Comm. Pure Appl. Math.}, 41(4):459--469, 1988.

\bibitem{Turok1990}
N.~Turok and D.~Spergel.
\newblock Global texture and the microwave background.
\newblock {\em Phys. Rev. Lett.}, 64:2736--2739, 1990.

\end{thebibliography}
\end{document}